\documentclass[preprint,onecolumn,nofootinbib]{revtex4}
\pdfoutput=1
\usepackage[colorlinks=true,linkcolor=blue,urlcolor=blue,filecolor=black,citecolor=red,pdfstartview=FitV,pdftitle={},pdfsubject={},pdfkeywords={},pdfpagemode=None,bookmarksopen=true]{hyperref}
\usepackage{graphicx}
\usepackage{amsmath}
\usepackage{amsfonts}
\usepackage{amssymb,ulem}
\usepackage{color}%
\usepackage{CJK}
\usepackage{dcolumn}
\newcommand{\f}{\begin{equation}}
\newcommand{\ff}{\end{equation}}
\newcommand{\fa}{\begin{eqnarray}}
\newcommand{\ffa}{\end{eqnarray}}

\begin{document}
\begin{CJK*}{GBK}{kai}
\title{ Weyl corrections to diffusion and chaos in holography}
\author{Wei-Jia Li $^{1}$}
\thanks{weijiali@dlut.edu.cn}
\author{Peng Liu $^{2}$}
\thanks{phylp@jnu.edu.cn}
\author{Jian-Pin Wu $^{3,4}$}
\thanks{jianpinwu@mail.bnu.edu.cn}
\affiliation{
$^1$ Institute of Theoretical Physics, School of Physics, Dalian
University of Technology, Dalian 116024, China\ \\
$^2$ Department of Physics, Jinan University, Guangzhou 510632, China\ \\
$^3$Center for Gravitation and Cosmology, College of Physical Science and Technology,
Yangzhou University, Yangzhou 225009, China\ \\
$^4$ Institute of Gravitation and Cosmology, Department of
Physics, School of Mathematics and Physics, Bohai University, Jinzhou 121013, China}
\begin{abstract}
Using holographic methods in the Einstein-Maxwell-dilaton-axion (EMDA)
theory, it was conjectured that the thermal diffusion in a strongly coupled metal without quasi-particles saturates an universal lower bound that is associated with the chaotic property of the system at infrared (IR) fixed points~\cite{blake:1705}. In this paper, we investigate the thermal transport and quantum chaos in the EMDA theory with a small Weyl coupling term. It is found that the Weyl coupling correct the thermal diffusion constant $D_Q$ and butterfly velocity $v_B$ in different ways, hence resulting in a modified relation between the two at IR fixed points. Unlike that in the EMDA case, our results show that the ratio $D_Q/ (v_B^2\tau_L)$ always contains a {\it non-universal} Weyl correction which depends also on the bulk fields as long as the $U(1)$ current is marginally relevant in the IR.

\end{abstract}
\maketitle
\tableofcontents
\section{Introduction}
Investigation of the thermoelectric transport in metallic systems is one of core topics in modern condensed matter physics. In contrast to the weakly coupled metals whose dynamics are governed by long-lived quasi-particles,  the transport properties of the strongly correlated metals with no single particle excitations are described by  the emergent hydrodynamic like degrees of freedom. Moreover, a wide class of such systems exhibit an universal Planckian relaxation timescale, $\tau_p\sim\hbar/(k_BT)$ (set $\hbar=k_B=1$) \cite{Sachdevbook,Hartnoll:2016apf}.

A well-known category of the strongly correlated metals are the so-called ``bad metals" or ``incoherent metals". In these systems, the resistivity increases linearly with temperature and violates the Mott-Ioffe-Regel (MIR) bound, and there is no sharp Drude peak in the AC conductivities at high temperatures due to rapid momentum relaxation. Because of the breakdown of the single particle approximation and other perturbative methods, these features still lack a deep understanding within the conventional QFT. Motivated by the observation that in incoherent metals the momentum dissipation depends heavily on the microscopic details of materials, which should not be the underlying reason of the universal strange metal, S. Hartnoll proposed that strange metals could be explained by the saturation of diffusion bounds $D_{c,Q}\gtrsim v_F^2/T$ where $v_F$ is the Fermi velocity \cite{hartnoll:2014}\footnote{Notice that diffusion bound is similar to and is partly motivated by the famous Kovtun-Son-Starinets (KSS) bound on the momentum diffusion that was found in the holographic studies of plasma \cite{kss:2004}}.
However, the Fermi velocity is in general not sharply defined in the systems without quasi-particles.

The holographic duality provides us an tractable approach to the physics with no quasi-particles. It has been widely applied to studying the transport properties of strongly correlated systems.  In holography, the DC conductivities can be captured by fluid like dynamics near the black hole horizon via the membrane paradigm \cite{liu:2009,donos:2014}. Based on the Einstein-Maxwell-dilaton-axion (EMDA) theories, M. Blake proposed a connection between the thermoelectric transport  and quantum chaos in strongly coupled systems that\cite{blake:1603,blake:1604}
\begin{eqnarray}\label{dbound}
D_{c,Q}= C_{c,Q} v_B^2\tau_L\,,
\end{eqnarray}
where $C_{c,Q}$ are constants only depending on the scaling properties of the IR fixed points, $v_B$ is the butterfly velocity characterizing the speed of information spreading, $\tau_L$ is the Lyapunov timescale characterizing the growth of the chaos which saturates its maximum $\frac{1}{2\pi T}\sim \tau_p$  in holographic systems and the Sachdev-Ye-Kitaev(SYK) models \cite{shenker:2013,malda:2015,gu:2016} but is much longer in quasi-particle systems \cite{Aleiner:2016,werman:2017}.  Then this bound seems valid for arbitrary chaotic systems with or without Fermi velocity. Whereas, it has been found that the bound on the charge diffusion can be violated in striped systems \cite{lucas:2016} or theories with higher derivative terms \cite{baggioli:2016}. Recently, it was pointed out that $v_B^2\tau_L$ may bound only the thermal diffusion instead of the charge diffusion with:
\begin{eqnarray}\label{conts}
C_{Q}=\frac{z}{2z-2} \ \ \text{at generic non-relativistic fixed points,}
\end{eqnarray}
where $z$ is the dynamical exponent \cite{blake:1705}. Then the ratio of $D_Q$ to $v_B^2\tau_L$ is quite universal, as it only depends on the scaling property of the IR theory, regardless of  the UV parameters of the matter fields, say, the chemical potenial/charge density, magnitude of the lattice, etc.

However, it is still unclear wether (\ref{conts}) universally holds or not in holography. The bottom-up approach allows us to touch this question in any (generalized) gravity theories with self-consistency. The bound (\ref{conts}) has been checked in many cases, and seems to work well in holography so far \cite{lucas:2016,baggioli:2016,baggioli:2017,blake:1611,kim:1704,kim:1708}\footnote{In a recent paper \cite{liu:2017}, it was reported that the diffusion bound can be violated in a higher derivative gravity theory. However, in these kinds of theories, there are two distinct butterfly velocities even in isotropic systems which seems quite odd from the angle of condensed matter physics.}.  Nevertheless, the condensed matter models studied in \cite{Patel:2016} and \cite{gu:1702} have already revealed two counter-examples. Then, it is worth exploring to what extent (\ref{conts}) holds in holography. Suppose the proposed universal $C_Q$ is somehow changed, it should be the two following situations:

a. $C_Q$ is still geometry-dependent only, but the relation (\ref{conts})  is modified due to certain pure gravity corrections.

b. $C_Q$ may also depend on the details of matter fields due to other kinds of corrections, which makes its expression totally non-universal.\\
Either case provides a necessary condition for the complete violation of the bound.

In this paper, we focus on the second one. A practicable way of modifying the holographic theory is to add the Weyl coupling terms, which couples the gauge field with the Weyl tensor. Previously, the effects of this kind of terms have been studied in a variety of holographic models \cite{Myers:2010pk,Wu:2010pk,Ma:2011zze,Zhao:2012kp,Zhang:2015eea,Momeni:2014efa,WitczakKrempa:2012gn,WitczakKrempa:2013ht,Witczak-Krempa:2013nua,Katz:2014rla,Sachdev:2011wg,Hartnoll:2016apf,Witczak-Krempa:2013aea,
Ritz:2008kh,Wu:2016jjd,Fu:2017,Dey:1510,Dey:1512,Yling:2017,Mahapatra:2016}. Here, we consider an EMDA action coupled with a small Weyl coupling term and investigate the Weyl corrections on the thermal diffusivity, the butterfly velocity and the ratio $C_Q$. The content of the paper is as
follows: In section \ref{HW}, we introduce the holographic action and the black hole
solutions. In section \ref{diffusion}, we analyze the thermal diffusion, butterfly velocity
and their  relation at low temperatures. In section \ref{cond}, we conclude. And the technical details are shown in the appendix.\\
\textbf{Note added:} As this work was being completed, \cite{Mokhtari:2017} appeared which has some overlap with our discussions.

\section{Holographic action and black holes}\label{HW}

We consider the four dimensional Einstein-Maxwell-dilaton theory coupled to two axionic scalars $\chi^I$ associated with the translational symmetry breaking and a Weyl  coupling term.
\begin{eqnarray}\label{haction}
&&
S=S_{\text{EMDA}}+S_{\text{Weyl}}\,,
\nonumber
\\
&&
S_{\text{EMDA}}=\int d^4x \sqrt{-g} \left(R-\frac{1}{2}(\partial \phi)^2-V(\phi)-\frac{1}{2}W(\phi)(\partial \chi^I)^2-\frac{1}{4}Z(\phi)F^2\right)
\,,
\nonumber
\\
&&
S_{\text{Weyl}}=  \gamma \int d^4x \sqrt{-g} \left( U(\phi)C_{\mu\nu\rho\sigma}F^{\mu\nu}F^{\rho\sigma}\right)
\,,
\label{a1}
\end{eqnarray}
with the indexes $I=x,y$ and the Weyl coupling $\gamma$. In the action above we have taken $16\pi G=L=1$ and Einstein's convention for convenience.
By definition, the Weyl tensor in four dimensions is given by
\begin{eqnarray}\label{wtensor}
C_{\mu\nu\rho\sigma}=R_{\mu\nu\rho\sigma}+\frac{1}{2}\left(g_{\mu\sigma}R_{\rho\nu}+g_{\nu\rho}R_{\mu\sigma}-g_{\mu\rho}R_{\sigma\nu}-g_{\nu\sigma}R_{\rho\mu}\right)+\frac{1}{6}
\left(g_{\mu\rho}g_{\nu\sigma}-g_{\mu\sigma}g_{\nu\rho}\right)R\,.
\end{eqnarray}

Adding Weyl couplings will, in general, bring about higher order differential equations which makes the problem mathematically difficult. So we will only consider the charged case with a small $\gamma$ coupling and expand the results up to the linear power in $\gamma$. The generic ansatz for isotropic solutions should be
\begin{eqnarray}\label{a4}
&ds^2=-f(r)dt^2+h(r)d r^2+g(r) (dx^2+dy^2)\,, \nonumber\\
&A_\mu=A_t(r), \ \ \chi^I=k \delta_{i}^Ix^i\,, \ \ i=x,y,
\end{eqnarray}
whose IR geometry can be classified into several distinct cases, depending on the couplings $U$, $V$, $W$ and $Z$.
\subsubsection*{Lifshitz/Hyperscaling violating geometries}
In the EMDA theory without Weyl corrections, the background solution can be Lifshitz/Hyperscaling violating geometry in the IR at low temperatures. This have been analyzed and classified into several different cases depending on the scaling properties in the IR \cite{Gouteraux:1401} . So here we just review this briefly.
These solutions can be achieved by setting the following exponential potentials
\begin{eqnarray}\label{epotential}
V(\phi)= - V_0 e^{-\delta \phi}\,, \ \ W(\phi)= e^{\lambda \phi}, \ \ Z(\phi)=e^{\eta \phi}\,,
\end{eqnarray}
which gives the near extremal IR solution
\begin{eqnarray}\label{blackf}
&&
f(r)=r^{\theta-2z}\left[1-\Big(\frac{r}{r_h}\Big)^{2+z-\theta}\right]\,,\,\,h(r)=L^2r^{\theta-2}\left[1-\left(\frac{r}{r_h}\right)^{2+z-\theta}\right]^{-1}\,,
\nonumber\\
&&
g(r)=\tilde{L}^2r^{\theta-2}\,,\,\, \phi=\varphi_0 \text{log}r\,,\,\, A_t(r)=a_0 r^{\zeta-z}\left[1-\Big(\frac{r}{r_h}\Big)^{2+z-\theta}\right]\,,
\end{eqnarray}
where $z$ and $\theta$ are dynamical and hyperscaling violating exponents respectively, ${\varphi}_0$ depends only on the scaling exponents $z$, $\theta$ and $\zeta$, while $L$, $\tilde{L}$, $a_0$ depend not only on the scaling exponents but also $V_0$ and the magnitude of the axionic lattice, $k$. In the extremal limit, the black hole solution flows towards different IR fixed points with the following features: \\
(a) Current $\&$ axion are both marginally relevant; (b) Current is marginally relevant $\&$ axion is irrelevant; (c) Current is irrelevant $\&$ axion is marginally relevant; (d) Current $\&$ axion are both irrelevant.

On top of that, we add the Weyl coupling and set $U(\phi)=e^{u \phi}$. Turning on a Weyl term may change the IR geometries/fixed points significantly. For simplicity, we can choose such values of $u$ that the Weyl corrections are at the same order in powers of the radial coordinate as the terms from the original Maxwell term.\footnote{In this paper, we only consider this special case and study the impacts of such a Weyl term on the thermal transport and chaos. A detailed analysis on how the general $RF^2$-like couplings affect the IR geometry will be presented in future work\cite{Li:17xx}.} Then one can show that the small $\gamma$ coupling just slightly changes the background geometry through modifying the parameters $a_0$, $\varphi_0$ and $L$ (See the details in appendix \ref{appB}). Nevertheless, the IR property should still be the Lifshitz/hyerscaling violating type.

\subsubsection*{$\text{AdS}_2 \times \text{R}^2$ geometries}
The black hole solution (\ref{metric2}) can also flow towards the $\text{AdS}_2 \times R^2$ fixed points in the IR.  In these cases, we have
\begin{eqnarray}\label{metric3}
f=R(r-r_e)^2, \ \ g=g_e, \ \ \phi=\phi_e,
\end{eqnarray}
where the constants $R$, $g_e$ and $\phi_e$ are constrained by
\begin{eqnarray}\label{constraintb2}
&2R\left(1+\frac{2\gamma q^2 U(\phi_e)}{3g_e^2Z(\phi_e)^2}\right)\approx\frac{k^2W(\phi_e)}{g_e}+\frac{q^2}{g_e^2Z(\phi_e)},\\
&0\approx2V(\phi_e)+\frac{2k^2W(\phi_e)}{g_e}+\frac{q^2}{g_e^2Z(\phi_e)}+O(\gamma^2),\\
&0\approx2V'(\phi_e)+\frac{2k^2W'(\phi_e)}{g_e}-\frac{q^2Z'(\phi_e)}{g_e^2Z(\phi_e)^2}-\frac{8\gamma R q^2(Z(\phi_e)U'(\phi_e))-2U(\phi_e)Z'(\phi_e)}{3g_e^2Z(\phi_e)^3}.
\end{eqnarray}
with the location of the extremal horizon at $r=r_e$ and $R$ is a dimensionless constant that depends on $\gamma$, the gauge field and axion at the horizon.
Turning on a small temperature, the black hole solution is slightly deformed as
\begin{eqnarray}\label{metric4}
f(r)=R\left[(r-r_e)^2-r_\epsilon^2\right]\,,
\end{eqnarray}
where $r_\epsilon$ is a small deviation from the extremal horizon and the external horizon is $r=r_h=r_e+r_\epsilon$.  Then $r_\epsilon=\frac{2\pi T}{R}$.
For the trivial case
\begin{eqnarray}\label{normal}
Z=W=U=1,
\end{eqnarray}
the full analytic solution has been found in \cite{Yling:2017}. It is
\begin{eqnarray}\label{metric2}
&&f(r)=f_0(r)+\gamma Y(r), \ h(r)=\frac{1}{f(r)}
\,,
\nonumber\\
&&g(r)=g_0(r)+\gamma G(r)=r^2+\gamma G(r)\,, \nonumber\\
&&A_t(r)=A_{t0}(r)+\gamma H(r)=\mu-\frac{q}{r}+\gamma H(r)\,,
\end{eqnarray}
where $f_0$ and $g_0$ are the metric without the Weyl correction, $\mu$ is the chemical potential, $q$ is the charge density, $G(r)=\frac{q^2}{9 r^2}$, $Y(r)$ and $H(r)$ are complicated functions of $q$, $k$ and $r$ whose forms are not important in our discussions. In the extremal limit, we have $f'(r_h)=0$ with $r_h\neq0$ as long as the current or/and the axion is/are non-vanishing. Then, the IR geometry should be $\text{AdS}_2 \times \text{R}^2$.

In this paper, we will  focus on the general $AdS_2\times R^2$ domain wall solution. The detailed IR analysis has been shown in appendix B2.
\section{Thermal diffusion and butterfly velocity}\label{diffusion}
For convenience, we introduce a new radial coordinate as in \cite{blake:1705,kim:1708}\footnote{As is found in \cite{kim:1708} that $z\neq\theta$. Therefore, the new coordinate is always well-defined.}
\begin{eqnarray}\label{newc}
\tilde{r}=\left|\frac{L}{\theta-z}\right| r^{\theta-z}\,.
\end{eqnarray}
Then, the background metric can be rewritten as
\begin{eqnarray}\label{newmetric}
&&ds^2=-f(\tilde{r})dt^2+f(\tilde{r})^{-1}d \tilde{r}^2+g(\tilde{r}) (dx^2+dy^2),\nonumber\\ && \ f(\tilde{r})=L_t^{-2} \tilde{r}^{\frac{2 z-\theta}{z-\theta}}\left[1-\left(\frac{\tilde{r}_h}{\tilde{r}}\right)^{\frac{2+z-\theta}{z-\theta}}\right], \ g(\tilde{r})=\bar{L}_x^{-2} \tilde{r}^{\frac{2-\theta}{z-\theta}}, \ \phi=\Phi_0 \text{log}\tilde{r}, \ A_t=A_0 \tilde{r}^{\frac{\zeta-z}{\theta-z}},
\end{eqnarray}
where
\begin{eqnarray}\label{newmetric2}
L_t^2=\left|\frac{L}{\theta-z}\right|^{\frac{2z-\theta}{z-\theta}}, \ L_x^2=\frac{1}{\tilde{L}^2}\left|\frac{L}{\theta-z}\right|^{\frac{\theta-2}{\theta-z}}, \ \Phi_0=\frac{\varphi_0}{z-\theta}, \ A_0=a_0\left|\frac{\theta-z}{L}\right|^{\frac{\zeta-z}{\theta-z}}.
\end{eqnarray}
Performing the Donos-Gauntlett strategy \cite{donos:2014}, we can express the DC conductivities just in terms of the metric components and $A_t'$ at the horizon (See the details in Appendix \ref{appC}.). Our result implies that the time-reversal symmetry is violated at $O(\gamma)$ when $A_t'\neq0$ according to the Onsager relation \cite{Onsager:1931}. Moreover, it has been revealed in \cite{lucas:2016,baggioli:1602,Gouteraux:1602,baggioli:2016} that the conjectured bounds on the electric conductivity $\sigma=1$ as well as that on the charge diffusion $D_c\sim v_B^2 \tau_L$ can both be violated in general holographic models. Therefore, in this work, we focus only on the thermal transport.

The open-circuit thermal conductivity at low temperatures is given by
\begin{eqnarray}\label{Tconducitivity}
\kappa=4 \pi\left(1-\frac{2}{3}\gamma U{A'_t}^2 \right)\frac{f'}{f''}\Big|_{\tilde{r}=\tilde{r}_h}+O(\gamma {f'}^2 |_{\tilde{r}=\tilde{r}_h})\,.
\end{eqnarray}
Now the prime refers to the derivative with respect to $\tilde{r}$. In contrast to that in Einstein gravity, it can never be expressed merely in terms of the near horizon geometry.\footnote{If one try to eliminate $A_t'$ by using the Einstein equation, the final result will also depend on $k^2W$, $Z$ and $U$.} Then the thermal diffusivity can be calculated via the following Einstein relation:
\begin{eqnarray}\label{Teinstein}
D_Q=\frac{\kappa}{c_q},
\end{eqnarray}
where $c_q$ is the heat capacity with fixed charge density which is defined as
\begin{equation}\label{capacity}
c_{q}\equiv T \frac{d s}{d T}\Big|_{q}\,.
\end{equation}
Following \cite{blake:1705}, we will compute $D_Q$ and compare it with the results of the butterfly velocity at the IR fixed points that we have discussed in the previous section.
\subsubsection*{Generic fixed points}
 The entropy density can be calculated by the {\it Wald formula} \cite{wald:1993,visser:1993,Brustein:2009}, which gives
\begin{eqnarray}\label{entropyd}
s=4 \pi g\left(1-\frac{2\gamma}{3}UA_t'^2\right)\Big|_{\tilde{r}=\tilde{r}_h}
\,.
\end{eqnarray}
Obviously, the factor $U(\tilde{r_h})A_t'(\tilde{r}_h)^2$, plays a crucial role of modifying the thermal diffusion in (\ref{Teinstein}) and the entropy density in (\ref{entropyd}), hence the heat capacity as well. In the small $\gamma$ expansions, we can just take the value of $U(\tilde{r_h})A_t'(\tilde{r}_h)^2$ in the EMDA theory.

When the current is {\it marginally relevant}, i.e, $\zeta=\theta-2$ and $\Phi_0 u=\frac{4}{\theta-z}$, one have $A_t(r)=A_0 \tilde{r}^{\frac{2+z-\theta}{z-\theta}}$.  Then we find that
\begin{eqnarray}\label{Aprime}
U(\tilde{r_h})A_t'(\tilde{r}_h)^2=\left(\frac{2+z-\theta}{z-\theta}\right)^2A_0^2,
\end{eqnarray}
which is temperature-independent. Then (\ref{Tconducitivity}) and (\ref{capacity}) can be rewritten as
\begin{eqnarray}\label{rtc}
&\kappa\equiv \mathcal{A} \kappa_0, \ c_q\equiv \mathcal{A}c_{q0}, \nonumber\\
 & \mathcal{A}=\left[1-\frac{2}{3}\gamma\left(\frac{2+z-\theta}{z-\theta}\right)^2A_0^2\right]
 \end{eqnarray}
where $\kappa_0$ and $c_{q0}$ represent the thermal conductivity and heat capacity obtained in the EMDA theory. Applying (\ref{newmetric}) and (\ref{Teinstein}), the thermal diffusion is obtained as
\begin{equation}\label{thermalc}
D_Q\approx \frac{z(z-\theta)}{2(2-\theta)(z-1)}L_x^2\tilde{r}_h^{\frac{z-2}{z-\theta}}
\end{equation}
which is not modified by the Weyl coupling. On the other hand, the butterfly velocity can be obtained by performing the shockwave calculations.  The details have been shown in appendix \ref {appxshock}. It can also expressed in terms of the horizon data
\begin{eqnarray}\label{butterflyv2}
v_B^2\tau_L&&\approx\frac{1}{g'}-\frac{2 \gamma U g {A_t'}^2 f''}{3 f' g'^2}+\frac{2 \gamma
  U A_t'^2}{g'}+\frac{4 \gamma  g U A_t' A_t''(r)}{3
   g'^2}\Big |_{\tilde{r}=\tilde{r}_h}
   \,,
   \nonumber\\
   &&=\frac{z-\theta}{2-\theta}\left[1-\frac{2\gamma(2z+3\theta-12)(2+z-\theta)^2A_0^2}{3(z-\theta)^2(2-\theta)}\right]L_x^2\tilde{r}_h^{\frac{z-2}{z-\theta}}
   \,.
   \end{eqnarray}
This further requires that $\theta\neq 2$. Finally, we obtain that the ratio of (\ref{thermalc}) to (\ref{butterflyv2}) is
\begin{eqnarray}\label{ratio1}
C_Q\equiv \frac{D_Q}{v_B^2\tau_L}\approx \frac{z}{2z-2}\left[1+\frac{2\gamma(2z+3\theta-12)(2+z-\theta)^2A_0^2}{3(z-\theta)^2(2-\theta)}\right].
\end{eqnarray}
at the generic fixed points when the current is marginally relevant in the IR. The interesting thing is that there is always a {\it non-universal} correction that comes from the Weyl corrections, as one can see from (\ref{newmetric2}), (\ref{IR1}) and (\ref{IR2}) that the constant $A_0$ highly depends on the details of the matter fields in the IR region.

While if the current is {\it irrelevant} and the axion is marginally relevant in the IR, the Weyl correction is vanishing in the extremal limit. Then, at this IR fixed point,
\begin{eqnarray}\label{ratio2}
C_Q=\frac{z}{2z-2}.
\end{eqnarray}

If the current and axion are {\it both irrelevant}, $z=1$. In this case $D_Q$ is controlled by an irrelevant deformation and $C_Q\gg1$, which is not universal even in Einstein gravity\cite{blake:1705}.

\subsection*{$\text{AdS}_2 \times \text{R}^2 $ fixed points}
For this class of geometries, $g=g_e$ is a constant. And, in contrast to the Lifshitz/hyperscaling violating cases, $c_q$ and $v_B$ should be determined by the leading irrelevant deformation of the fixed point solution.  Expanding $g(r)$ around its extremal value,  we obtain
\begin{eqnarray}\label{Ltfinal}
g(r)=g_e+\delta g_1+...=g_e+c_1 (r-r_e)^{1+\alpha \gamma}+...,
\end{eqnarray}
where $c_1$ is a constant that is fixed by the UV data and $\alpha$ is a parameter whose form has been shown explicitly in appendix B2.  In general, $\delta g_1$  contains two modes of dimensions $\Delta=2+\alpha \gamma$ and $\Delta_\phi$. To have a well-defined small $\gamma$ expansion, we should require that $\Delta_\phi>2+\alpha$. The details can be seen in appendix B2. Then, the expression (\ref{Ltfinal}) captures the leading behavior of $\delta g$. As a result, we have $g'(r_h)=c_1 (1+\alpha \gamma) r_\epsilon^{\alpha \gamma}=c_1 (1+\alpha \gamma) \left(\frac{2\pi T}{R}\right)^{\alpha \gamma}$.

The thermal conductivity and the entropy density can be written as
\begin{eqnarray}\label{rtc2}
&\kappa\equiv 4\pi\mathcal{B} \frac{f'(r_h)}{f''(r_h)}, \ s\equiv \mathcal{B}s_{0}= 4\pi\mathcal{B}g(r_h), \nonumber\\
  &\mathcal{B}=1-\frac{2\gamma U(\phi_h)q^2}{3Z(\phi_h)^2r_h^4}.
\end{eqnarray}
Then the entropy density is
\begin{eqnarray}\label{edensity2}
s=s_e+4\pi\mathcal{B}c_1\left(\frac{2\pi T}{R}\right)^{1+\alpha \gamma}+...\,,
\end{eqnarray}
where $s_e$ is the extremal entropy. And the thermal diffusion is given by
\begin{eqnarray}
D_Q\approx \frac{R^{\alpha\gamma}}{c_1 (1+\alpha\gamma)(2\pi T)^{\alpha \gamma}}\,.
\end{eqnarray}
At low temperatures, we have $f''(r_h)\gg f'(r_h)\rightarrow 0$. Then the second term in (\ref{butterflyv}) dominates over the other two Weyl correction terms. And the butterfly velocity can be written as\footnote{Through out our discussions, we always do the small $\gamma$ expansion before taking the low temperature limit.}
\begin{eqnarray}
v_B^2\tau_L&=&\frac{1}{ g'}-\frac{2\gamma U f'' g A_t'^2}{3 f'{g'}^2}\Big|_{r=r_h}+...
\,,
\nonumber\\
&\approx&\frac{R^{\alpha\gamma}}{c_1 (1+\alpha\gamma)(2\pi T)^{\alpha \gamma}}\left(1-\frac{2\gamma R_0 U(\phi_h) q^2}{3c_1^0\pi Z(\phi_h)^2 g_eT}\right)
\end{eqnarray}
where $R_0=\frac{k^2W(\phi_e)}{2g_e}+\frac{q^2}{2g_e^2Z(\phi_e)}$, $c_1^0=c_1(\gamma=0)$ and $g_e=r_e^2$. We thereby achieve that
\begin{eqnarray}
C_Q\approx1+\frac{2\gamma R_0 U(\phi_h) \mu^2}{3c_1^0\pi Z(\phi_h)^2 T}.
\end{eqnarray}
for $\gamma\ll\frac{T}{\mu}\ll1$ while fixing the other quantities. Finally, we find that there is again a {\it non-universal} correction for the finite density case.
\section{Conclusion and discussion}\label{cond}
In this paper, we have studied the thermal transport and butterfly effects by performing the holographic calculations in the EMDA model coupled with a small Weyl coupling term. It is found that the ratio of thermal diffusion $D_Q$ to the butterfly velocity times the Lyapunov timescale $v_B \tau_L$ contains a non-universal Weyl correction when the Weyl coupling terms are marginally relevant in the IR.

When the IR geometry is Lifshitz or hyperscaling violating type, the form of $D_Q$ remains unchanged while the butterfly velocity can get corrected. Then, the Weyl correction in $C_Q$ depends not only on the scaling properties of the IR fixed point but also on the parameter of the gauge field $A_0$ and the Weyl coupling $\gamma$. When the IR geometry is $\text{AdS}_2\times\text{R}^2$, both of the thermal diffusion and the butterfly velocity can be modified. And the non-universal part in $C_Q$ can be explicitly expressed in terms of
the $\gamma$ and the UV paramters, $\mu$, $k$, etc. In both cases, the conjectured universal bound on $C_Q$ can be ``slightly violated" due to the Weyl corrections.  While, in the ``incoherent limit" \cite{davison:1411,davison:1505} which implies that $T$ is finite and the value of $k$ is far bigger than $T$ and any other parameters of the matter fields in the IR, we can just simply neglect the effect of $A_0$ or $\mu$ in the IR.  The Weyl corrections in $C_Q$ is thus vanishing.  This suggests that the proposed diffusion bound in \cite{blake:1705} could be valid only in the incoherent limit.
\begin{acknowledgments}

We are particularly grateful to M. Baggioli and  for valuable comments about the manuscript. We also would like to thank K.-Y. Kim and H.-S. Liu for their stimulating discussions. This work is supported by the Natural Science Foundation of China under
Grant Nos. 11375026, 11575195, 11775036 and 11305018.
W.-J. Li is also supported by the Fundamental Research Funds for the Central Universities No. DUT 16 RC(3)097.
J.-P. Wu is also supported by Natural Science Foundation of Liaoning Province under Grant No.201602013.

\end{acknowledgments}

\begin{appendix}

\section{Covariant form of the equations of motion}
The equations of motion from the holographic action (\ref{haction}) are given by
\begin{eqnarray}\label{a3}
&&
\nabla_{\mu}\left(Z(\phi) F^{\mu\nu}-4\gamma U(\phi)C^{\mu\nu\rho\sigma}F_{\rho\sigma}\right)=0\,,
\label{EE}
\\
&&
\nabla^2\phi-V'(\phi)-\frac{Z'(\phi)}{4} F^2+\gamma U' (\phi)C_{\mu\nu\rho\sigma}F^{\mu\nu}F^{\rho\sigma}-\frac{W'(\phi)}{2}(\partial \chi^I)^2=0\,,
\label{phiE}
\\
&&
\nabla_\mu(W(\phi) \nabla^\mu \chi^I)=0\,,
\label{phiE}
\\
&&
R_{\mu\nu}-\frac{1}{2}R g_{\mu\nu}+\frac{(\partial\phi)^2}{4}g_{\mu\nu}+\frac{V(\phi)}{2}g_{\mu\nu}
-\frac{Z(\phi)}{2}\Big(F_{\mu\rho}F_{\nu}^{\ \rho}-\frac{1}{4}g_{\mu\nu}F_{\rho\sigma}F^{\rho\sigma}\Big)
-\frac{1}{2}\partial_\mu\phi\partial_\nu\phi
\nonumber
\\
&&
-\frac{W(\phi)}{2}\Big(\partial_\mu\chi^I\partial_\nu\chi^I-\frac{g_{\mu\nu}}{2}(\partial \chi^I)^2\Big)-\gamma U(\phi)\Big(G_{1\mu\nu}+G_{2\mu\nu}+G_{3\mu\nu}\Big)
=0\,,
\label{EE}
\end{eqnarray}
with the Weyl corrections:
\begin{eqnarray}\label{GG}
G_{1\mu\nu}&=&\frac{1}{2}g_{\mu\nu}R_{\alpha\beta\rho\sigma}F^{\alpha\beta}F^{\rho\sigma}
-3R_{(\mu|\alpha\beta\lambda|}F_{\nu)}^{\ \alpha}F^{\beta\lambda}
-2\nabla_{\alpha}\nabla_{\beta}(F^{\alpha}_{\ (\mu}F^{\beta}_{\ \nu)})
\,,
\nonumber
\\
\nonumber
G_{2\mu\nu}&=&-g_{\mu\nu}R_{\alpha\beta}F^{\alpha\lambda}F^{\beta}_{\ \lambda}
+g_{\mu\nu}\nabla_{\alpha}\nabla_{\beta}(F^{\alpha}_{\ \lambda}F^{\beta\lambda})
+\Box(F_{\mu}^{\ \lambda}F_{\nu\lambda})
-2\nabla_\alpha\nabla_{(\mu}(F_{\nu)\beta}F^{\alpha\beta})
\nonumber
\\
&&
+2R_{\nu\alpha}F_{\mu}^{\ \beta}F^{\alpha}_{\ \beta}
+2R_{\alpha\beta}F^{\alpha}_{\ \mu}F^{\beta}_{\ \nu}
+2R_{\alpha\mu}F^{\alpha\beta}F_{\nu\beta}
\,,
\nonumber
\\
\
G_{3\mu\nu}&=&
\frac{1}{6}g_{\mu\nu}RF^2-\frac{1}{3}R_{\mu\nu}F^2-\frac{2}{3}RF^{\alpha}_{\ \mu}F_{\alpha\nu}
+\frac{1}{3}\nabla_{(\mu}\nabla_{\nu)} F^2-\frac{1}{3}g_{\mu\nu}\Box F^2\,.
\end{eqnarray}
where the Laplacian is defined by $\Box=\nabla_\mu\nabla^\mu$.
\section{Analysis of the IR geometries}\label{appB}
\section*{B1. Hyperscaling violating geometries}
In the extremal limit, the IR solution (\ref{blackf}) reduces to
\begin{eqnarray}\label{blackfex}
f(r)=r^{\theta-2z}, \ h(r)=L^2r^{\theta-2}, \ g(r)=\tilde{L}^2r^{\theta-2}, \ \phi=\varphi_0 \text{log}r, \ A_t(r)=a_0 r^{\zeta-z}.
\end{eqnarray}
Plugging this into the Einstein equation, dilaton equation as well as Maxwell equation, one obtains that

\begin{eqnarray}\label{eomsex}\label{ee1}
&&6 L^4 V_0 r^{\theta -\delta  \phi_0}-\frac{6 k^2 L^4 r^{2+\lambda \phi_0}}{\tilde{L}^2}-3 L^2 \left(a_0^2 (z-\zeta )^2 r^{2 \zeta +\eta  \phi_0-\theta }+(\theta
   -8) \theta +\phi_0^2+12\right)\nonumber\\
&&+8 a_0^2 \gamma  (z-\zeta )^2\left((2 \zeta -\theta -3) (2 \zeta -\theta -2)-2z (z-1)\right)
   r^{2 \zeta -2 \theta +u \phi_0}
=0,\\ \label{ee2}
&&3 L^2 r^{-\theta -2} \left(a_0^2 (z-\zeta )^2 r^{2 \zeta +\eta  \text{$\phi
   $0}}+r^{\theta } \left((\theta -2) (3 \theta -4 z-2)-\phi_0^2\right)\right)+8  \gamma a_0^2 (z-1) (z-\zeta )^2\nonumber\\
 &&(2 (\zeta +z-1)-\theta ) r^{2 \zeta -2 \theta +u
   \phi_0-2}+6 L^4 \left(\frac{k^2 r^{\lambda  \phi_0}}{\tilde{L}^2}-V_0 r^{-\delta
   \phi_0+\theta -2}\right)=0,\\ \label{ee3}
 &&3 L^2 r^{\delta  \varphi_0 +\theta } \left[r^{\theta } \left(\theta ^2-4 \theta +\varphi_0 ^2-4 \theta  z+4 z (z+1)+4\right)-a_0^2 (z-\zeta )^2
   r^{2 \zeta +\varphi_0\eta}\right]+4\gamma a_0^2   (z-\zeta )^2\nonumber\\&&[(\theta +2-2 \zeta)^2-2 z^2+z (-2 \zeta +\theta +4)] r^{2\zeta +\varphi_0  (\delta +u)}-6 L^4 V_0 r^{3 \theta }=0,\\ \label{ee4}
&&\frac{a_0^2 (z-\zeta )^2 r^{2 \zeta -3 \theta } \left(3 \eta  L^2 r^{\varphi_0\eta   +\theta }+8 \gamma  u (z-1) z r^{\varphi_0  u}\right)}{6
   L^4}+r^{-\theta } \Big[\frac{\varphi_0  (\theta -z-2)}{L^2}-\frac{k^2 \lambda  r^{\varphi_0  \lambda +2}}{\tilde{L}^2}\nonumber\\
&&-V_0\delta r^{\theta -\delta  \varphi_0
   }\Big]=0, \\  \label{ee5}
&&3 L^2 (\zeta +\varphi_0\eta -2) r^{\varphi_0\eta +\theta }+8 \gamma  (z-1) z  (\zeta -\theta +\varphi_0  u-2)r^{\varphi_0  u}=0.
\end{eqnarray}
From now on, we assume that the Weyl corrections are at the same order in powers of the radial coordinate as the original Maxwell term. With (\ref{ee1})-(\ref{ee5}), following the analysis in \cite{Gouteraux:1401}, we conclude that

(a) Current $\&$ axion are both Marginally relevant:\\
In this case, $\theta$ and $z$ are not fixed, while $\zeta=\theta-2$, $\varphi_0 \lambda=-2$, $\varphi_0 \delta=\theta$, $\eta=-\delta-2\lambda$, $\varphi_0 u=4$ and
\begin{eqnarray}\label{IR1}
L^2&\approx&\frac{2 (1+z-\theta) (2+z-\theta)}{2 V_0-k^2}\nonumber\\
&+&\frac{2 \gamma  \left(288-\theta ^3+18 \theta ^2-120 \theta +4 z^3+8 z^2+\left(\theta ^2-48\right) z\right) \left(k^2 (\theta -2 z)+2 V_0
   (z-1)\right)}{3 \left(2 V_0-k^2\right) (1+z-\theta) (2 z+4-\theta)},\nonumber\\
\tilde{L}^2&=&1, \nonumber\\
\varphi_0^2&\approx& \theta ^2-2 (\theta -2) z-4\nonumber\\
&+&\frac{4 \gamma \left(k^2 (\theta -2 z)+2
   V_0 (z-1)\right) \left[-(\theta -6) ((\theta -12) \theta +40)+2 (\theta -2) z^2+(\theta -8) (\theta -2) z\right]}{3 (-\theta +z+1) (-\theta +2 z+4)},\nonumber\\
a_0^2&\approx&\frac{2 k^2 (2 z-\theta )-4V_0 (z-1)}{\left(k^2-2 V_0\right)
   (2+z-\theta)}\nonumber\\
&+&4 \gamma  \left(k^2 (\theta -2 z)+2 V_0 (z-1)\right)\Big[\frac{V_0 \left(-3 \theta ^2+\theta  (z+38)+6 (z-2) z-120\right)}{3 \left(k^2-2 V_0\right) (-\theta +z+1) (-\theta +z+2)^2}\nonumber\\
   &+&\frac{2 k^2 \left(-(\theta -6) ((\theta -11) \theta +32)-2 z^3+(\theta +2) z^2+(\theta -4) (2 \theta -15) z\right)}{3 \left(k^2-2 V_0\right)
   (-\theta +z+1) (-\theta +z+2)^2 (-\theta +2 z+4)}\Big].
 \end{eqnarray}
To obtain above expressions, we have used the small $\gamma$ expansion.

(b) Current is marginally relevant $\&$ axion is irrelevant:\\
In this case, $\theta$, $z$ and $\varphi_0 \lambda$ are not fixed, while $\zeta=\theta-2$, $\varphi_0 \eta=4-\theta$, $\varphi_0 \delta=\theta$, $\varphi_0 u=4$ and
\begin{eqnarray}\label{IR2}
L^2&\approx&\frac{(1+z-\theta) (2+z-\theta)}{V_0}\nonumber\\
&+&\frac{2 \gamma  (z-1) \left(-\theta ^3+18 \theta ^2-120 \theta +4 z^3+8 z^2+\left(\theta ^2-48\right) z+288\right)}{3 (1+z-\theta) (4+2z-\theta)},\nonumber\\
\tilde{L}^2&=&1, \nonumber\\
\varphi_0^2&\approx& \theta ^2-2 (\theta -2) z-4\nonumber\\
&+&\frac{8 \gamma
   V_0 (z-1)\left[-(\theta -6) ((\theta -12) \theta +40)+2 (\theta -2) z^2+(\theta -8) (\theta -2) z\right] }{3 (-\theta +z+1) (-\theta +2 z+4)},\nonumber\\
a_0^2&\approx&\frac{2(z-1)}{2+z-\theta}-\frac{4 \gamma  V_0 (z-1) \left[-3 \theta ^2+\theta  (z+38)+6 (z-2) z-120\right]}{3 (-\theta +z+1) (-\theta +z+2)^2}.
 \end{eqnarray}

(c) Current is irrelevant $\&$ axion is marginally relevant:\\
In this case, $\theta$, $z$, $\zeta$, $\varphi_0 \eta$ and $\varphi_0 u$ are not fixed, but $\varphi_0 \lambda=-2$, $\varphi_0\delta=\theta$ and
\begin{eqnarray}\label{IR3}
L^2&\approx&\frac{(1+z-\theta) (2+z-\theta)}{V_0},\nonumber\\
\tilde{L}^2&=&\frac{k^2(2z-\theta)}{2V_0(z-1)}, \nonumber\\
\varphi_0^2&\approx& \theta ^2-2 (\theta -2) z-4.
 \end{eqnarray}

(d) Current $\&$ axion are both irrelevant:\\
In this case $\zeta$, $\varphi_0 \lambda$, $\varphi_0 \eta$ and $\varphi_0 u$ are not fixed, while $z=1$, $\varphi_0 \delta=\theta$ and
\begin{eqnarray}\label{IR4}
L^2&\approx&\frac{(2-\theta) (3-\theta)}{V_0},\nonumber\\
\tilde{L}^2&=&1, \nonumber\\
\varphi_0^2&\approx& \theta ^2-2 \theta.
 \end{eqnarray}
 Obviously, only when the current is marginally relevant the Weyl coupling affects the background through correcting the parameters $L$, $\varphi_0$ and $a_0$.
Note that the poles $2z+4-\theta=0$, $1+z-\theta=0$ and $2+z-\theta=0$ in the above equations should be excluded.

Before closing this subsection, we present some comments on the stability of the Hyperscaling violating geometry with Weyl term.
To implement the mode analysis on our model. Specifically, we turned on the following mode expansion based on \eqref{blackfex}:
\fa
&&
  f(r) = r^{\theta-2z}\left( 1+ c_{1} r^{\beta} \right),\quad h(r) = L^{2} r^{\theta-2}\left(1+ c_{2}r^{\beta}\right), \quad g(r) = \tilde{L}^{2} r^{\theta -2} \left(1+ c_{3} r^{\beta}\right)\,,
  \nonumber
  \\
  &&
  \phi(r) = \varphi_{0} \log\left(r\left(1+c_{4} r^{\beta}\right)\right),\quad A_{t}(r) = a_{0} r^{\zeta - z} \left(1+ c_{5} r^{\beta}\right)\,.
  \label{HV-mode}
\ffa
  By inserting the above expansions into the equations of motion and extracting the linear part in $c_{i}$, we obtain
  \fa M_{ij} c_{j} =0\,, \ffa
  where each element of the matrix $M$ is a function of $\beta$ and other parameters of the system. In our model we find the matrix $M$ is too complicated to obtain an analytical solution for $\beta$ so as to study the staiblity of our system. However, we explored the $\det \left(M\right)$ and found that the solutions $\beta$ from $\det \left(M\right) =0$ will only contain terms at $0$-th and $1$-st order of $\gamma$, and hence the stability from mode analysis will not receive significant modification.

 \section*{B2. The $AdS_2 \times R^2$ domain wall solution}
In this case, the background geometry near the extremal horizon $r_e$ can be expressed as
\begin{eqnarray}\label{metricb2}
\bar{f}=R(r-r_e)^2, \ \ \bar{g}=g_e, \ \ \bar{\phi}=\phi_e,
\end{eqnarray}
where the constants $R$, $g_e$ and $\phi_e$ are constrained by
\begin{eqnarray}\label{constraintb2}
&2R\left(1+\frac{2\gamma q^2 U(\phi_e)}{3g_e^2Z(\phi_e)^2}\right)\approx\frac{k^2W(\phi_e)}{g_e}+\frac{q^2}{g_e^2Z(\phi_e)},\\
&0\approx2V(\phi_e)+\frac{2k^2W(\phi_e)}{g_e}+\frac{q^2}{g_e^2Z(\phi_e)}+O(\gamma^2),\\
&0\approx2V'(\phi_e)+\frac{2k^2W'(\phi_e)}{g_e}-\frac{q^2Z'(\phi_e)}{g_e^2Z(\phi_e)^2}-\frac{8\gamma R q^2(Z(\phi_e)U'(\phi_e))-2U(\phi_e)Z'(\phi_e)}{3g_e^2Z(\phi_e)^3}.
\end{eqnarray}
To have a small temperature, we can generalize the extremal solution by introducing a small deviation $r_\epsilon$ as follows,
\begin{eqnarray}\label{metricb2}
\bar{f}=R\left[(r-r_e)^2-r_\epsilon^2\right], \ \ \bar{g}=g_e, \ \ \bar{\phi}=\phi_e.
\end{eqnarray}
For simplicity, one can choose the  coordinates properly so that $r_e=0$ and the location of the horizon is located at $r_h=r_\epsilon$. Next, we need to add irrelevant modes that will connect the IR geometry back to the UV $AdS_4$ boundary. Perturb the black hole solution as
\begin{eqnarray}\label{pertb}
f=\bar{f}+\delta f,\\
g=\bar{g}+\delta g,\\
\phi=\bar{\phi}+\delta \phi,
\end{eqnarray}
where $\delta f$, $\delta g$ and $\delta \phi$ represent the small fluctuations. To obtain the heat capacity, we need to extract the leading behavior of $\delta g$ in the low temperature limit. At the linearized order, the eoms of  $\delta g_1$ and $\delta \phi_1$ are given by
\begin{eqnarray}\label{eom2b11}
 &&(\bar{f}\delta \phi_1')'-\left(V''(\phi_e)+\frac{k^2W''(\phi_e)}{g_e}-\frac{Z''(\phi_e) q ^2}{2 g_e^2
   Z(\phi_e)^2}+\frac{Z'(\phi_e)^2 q ^2}{g_e^2 Z(\phi_e)^3}\right)\delta \phi_1+\left(\frac{k^2W'(\phi_e)}{g_e^2}-\frac{Z'(\phi_e) q ^2}{g_e^3 Z(\phi_e)^2}\right)\delta g_1\nonumber\\
   &&+\frac{2\gamma q^2}{3g_e^3Z(\phi_e)^4}\Big[2 R g_e \Big(U''(\phi_e) Z(\phi_e)^2-2 Z''(\phi_e) U(\phi_e)
   Z(\phi_e)-4 U'(\phi_e) Z'(\phi_e) Z(\phi_e)\nonumber\\
   &&+6 Z'(\phi_e)^2 U(\phi_e)\Big) \delta\phi_1-4 R Z(\phi_e) \Big(U'(\phi_e)
   Z(\phi_e)-2 Z'(\phi_e) U(\phi_e)\Big)\delta g_1
   +Z(\phi_e)\Big(2 Z'(\phi_e) U(\phi_e)\nonumber\\&&-U'(\phi_e) Z(\phi_e)\Big) \left(\bar{f}'
   \delta g_1'-g_e \delta f_1''\right)\Big]=0\,,
\\
   \nonumber\\
 &&\frac{\delta g_1}{r}-\delta g_1'-\frac{2 \gamma q ^2 U'(\phi_e)  \delta\phi_1}{3 g_e r Z(\phi_e)^2}-\frac{4 \gamma  q ^2 Z'(\phi_e) U(\phi_e)
   \delta\phi_1'}{3 g_e Z(\phi_e)^3}+\frac{4 \gamma q ^2 Z'(\phi_e)  U(\phi_e) \delta\phi_1}{3 g_e r
   Z(\phi_e)^3}-\nonumber\\ &&\frac{2 \gamma  q ^2 U(\phi_e) \delta g_1'}{3 g_e^2 Z(\phi_e)^2}+
  \frac{2 \gamma  q ^2 U(\phi_e)
   \delta g_1}{3 g_e^2 r Z(\phi_e)^2}=0\,.
    \label{eom2b12}
\end{eqnarray}
These two  equations are rather complicated. However, since $\gamma$ is small, one can in principal replace the $\delta g_1'$ and $\delta f_1''$ in the Weyl correction terms with $\delta g_1$ and $\phi_1$  through the zero order relations. For the $\gamma=0$ case, we have\cite{blake:1611}
\begin{eqnarray}\label{eom2b10}
&&\delta g_1'=\frac{\delta g_1}{r},\\
 &&\delta f_1''=\left(\frac{k^2W'(\phi_e)}{g_e}-\frac{Z'(\phi_e) q ^2}{g_e^2 Z(\phi_e)^2}\right)\delta \phi_1-\left(\frac{2R}{g_e}+\frac{q ^2}{g_e^3 Z(\phi_e)}\right)\delta g_1.
\end{eqnarray}
We can thereby simplify (\ref{eom2b11}) and (\ref{eom2b12}) by inserting the above relations into the Weyl terms. This gives
\begin{eqnarray}\label{eom2b2}
(\bar{f}\delta \phi_1')'-R\Delta_0(\Delta_0-1)\delta \phi_1+\mathcal{F}\delta g_1=0,
\\
\frac{\delta g_1}{r}-\delta g_1'+\gamma \mathcal{G}\frac{\delta \phi_1}{r}-\gamma \mathcal{H}\delta \phi_1' =0,\label{eom2b3p}
\end{eqnarray}
where
\begin{eqnarray}\label{eom2b3}
R\Delta_0(\Delta_0-1)&&=V''(\phi_e)+\frac{k^2W''(\phi_e)}{g_e}-\frac{Z''(\phi_e) q ^2}{2 g_e^2
   Z(\phi_e)^2}+\frac{Z'(\phi_e)^2 q ^2}{g_e^2 Z(\phi_e)^3}+\frac{2\gamma q^2}{3g_e^3Z(\phi_e)^4}\nonumber\\
  && \Big[2 R g_e \Big(U''(\phi_e) Z(\phi_e)^2-2 Z''(\phi_e) U(\phi_e)
   Z(\phi_e)-4 U'(\phi_e) Z'(\phi_e) Z(\phi_e)
  \nonumber\\
  &&+6 Z'(\phi_e)^2 U(\phi_e)\Big) -Z(\phi_e)\Big(2 Z'(\phi_e) U(\phi_e)-U'(\phi_e) Z(\phi_e)\Big)\nonumber\\
  &&\Big(k^2W'(\phi_e)-\frac{Z'(\phi_e) q ^2}{g_e Z(\phi_e)^2}\Big)\Big], \nonumber
  \end{eqnarray}
\begin{eqnarray}
 \mathcal{F}&&=\frac{k^2W'(\phi_e)}{g_e^2}-\frac{Z'(\phi_e) q ^2}{g_e^3 Z(\phi_e)^2}+\frac{2\gamma q^2}{3g_e^3Z(\phi_e)^4}\Big[Z(\phi_e)\Big(2 Z'(\phi_e) U(\phi_e)-U'(\phi_e) Z(\phi_e)\Big)\nonumber\\
  &&\Big(4R+\frac{q ^2}{g_e^2 Z(\phi_e)}\Big)-4 R Z(\phi_e) \Big(U'(\phi_e)
   Z(\phi_e)-2 Z'(\phi_e) U(\phi_e)\Big)\Big],\nonumber\\
  \mathcal{G}&&=\frac{4 \gamma q ^2  Z'(\phi_e) U(\phi_e) }{3 g_e
   Z(\phi_e)^3}-\frac{2 \gamma q ^2  U'(\phi_e)}{3 g_e Z(\phi_e)^2},\nonumber\\
   \mathcal{H}&&= \frac{4 \gamma q ^2 Z'(\phi_e)  U(\phi_e)}{3 g_e Z(\phi_e)^3}
\end{eqnarray}
Then, the eoms become two coupled first order differential equations.  Solving (\ref{eom2b2}) and (\ref{eom2b3p}) in the extremal limit, we obtain the following general solutions:
\begin{eqnarray}\label{soutions}
&&\delta g_1=c_1 r^{1+\alpha \gamma}+c_2 \gamma r^{\Delta_0-1+\beta \gamma},\\
&&\delta \phi_1= \frac{\mathcal{F} c_1}{R(\Delta_0^2-\Delta_0-2-3\alpha \gamma)} r^{1+\alpha \gamma}+\frac{\mathcal{F} c_2}{R\beta(1-2\Delta_0)} r^{\Delta_0-1+\beta \gamma},\label{soutions2}
\end{eqnarray}
where $\alpha=\frac{\mathcal{F}(\mathcal{G}-\mathcal{H})}{R(\Delta_0-2)(\Delta_0+1)}$, $\beta=\frac{\mathcal{F}\mathcal{H}(\Delta_0-1)-\mathcal{F}\mathcal{G}}{R(2\Delta_0-1)(\Delta_0-2)}$,  $c_1$ and $c_2$ are two integration constants which can be fixed by the UV data of the domain wall.  We find that there are two modes of dimensions $\Delta=2+\alpha \gamma$ and $\Delta_\phi=\Delta_0+\beta \gamma$ in both of (\ref{soutions}) and (\ref{soutions2}).

To achieve the solution above, we have assumed $\beta\neq0$. While for the $\beta=0$ case, one can easily check that the constant $c_2$ should be vanishing so that the second mode in (\ref{soutions2}) is regular. And there exist only one mode in $\delta g_1$.  Then, the solution is given by
\begin{eqnarray}\label{soutionsa}
&&\delta g_1=c_1 r^{1+\alpha \gamma},\\
&&\delta \phi_1= \frac{\mathcal{F} c_1}{R(\Delta_0^2-\Delta_0-2-3\alpha \gamma)} r^{1+\alpha \gamma}+\tilde{c}_2 r^{\Delta_0-1},\label{soutions2a}
\end{eqnarray}
For a general domain wall solution, $g$ and $\phi$ should be taken the form
\begin{eqnarray}\label{expansions}
&&g=\sum_{n_1,n_2\geq0}C_{n_1,n_2}^g r^{(1+\alpha \gamma)n_1+(\Delta_\phi-1)n_2},\\
&&\phi= \sum_{n_1,n_2\geq0}C_{n_1,n_2}^\phi r^{(1+\alpha \gamma)n_1+(\Delta_\phi-1)n_2},\label{expansions2}
\end{eqnarray}
where $n_1$ and $n_2$ are integers.\footnote{This form should be modified a little bit for the near $AdS_2$ case\cite{blake:1611}.}  We are interested in the leading correction to the extremal value of $g$ when we take $r=r_h\rightarrow0$ in (\ref{expansions}).

When $\gamma=0$, $\delta g_1$ only contains
an universal mode $\sim r$ with $n_1=1$ and $n_2=0$. Then, it is easy to see that the term with $n_1=0$ and $n_2=2$ will dominate over this universal mode  if $\Delta_\phi<3/2$. Therefore, the second order piece $\delta g_2$ gives the leading corrections in $g$.

However, with the Weyl corrections, the situation becomes subtle. When $\beta=0$, since $\delta g_1$ only contains the $r^{1+\alpha \gamma}$ mode,  $\delta g_1$ always gives the leading correction to the  extremal horizon if $\Delta_\phi>(3+\alpha \gamma)/2$. When $\beta\neq 0$,   $\delta g_1$ has two modes as is shown in (\ref{soutions}). Then, its leading behavior depends on wether the value of $\Delta_\phi$ is greater than $2+\alpha \gamma$. Regardless of the situation,  one can check that  the modes of $\delta g_2$ can never dominate over $\delta g_1$. As a result, $\delta g_1$ always supports the leading contribution in $\delta g$. Nevertheless, we should require $\Delta_\phi>2+\alpha \gamma$. Otherwise, the second mode in (\ref{soutions}) will dominate the behavior of $\delta g_1$ near the horizon. This is incompatible with the small $\gamma$ trick that we have used to simplify the original eoms (\ref{eom2b11}) and (\ref{eom2b12}).  In conclusion, we restrict $\Delta_\phi>2+\alpha \gamma$ so that $g$ can always be expanded as in (\ref{Ltfinal}).

\section{Derivation of the DC conductivities}\label{appC}
In order to calculate the DC conductivities, we introduce the following perturbations around the background
\begin{eqnarray}\label{a7}
&\delta A_x=(\zeta A_t(r)-E_x) t+a_x(r),\nonumber\\
&\delta g_{tx}=-\zeta f(r) t+g(r)h_{tx}(r),\nonumber\\
&\delta g_{rx}=g(r)h_{rx}(r),\nonumber\\
&\delta\chi^I=\psi^x(r).
\end{eqnarray}
where $f$, $g$, $A_t$ are the background fields in (\ref{newmetric}) and we have omitted the tilde symbol for the radial coordinate $\tilde{r}$ for simplicity.  All the linearized eoms can then be obtained by applying the ansatz ({\ref{a7}})
to (\ref{a3})-(\ref{GG}).

From the equation of motion of $a_x$, we can define a conserved current along the radial direction in the bulk:
\begin{eqnarray}\label{conservedcurrent1}
J^E&=&-f Z
   a_x' -g Z  A_t' h_{tx}-\frac{2 \gamma U  f^2 g'' a_x'}{3 g}+\frac{2 \gamma U  f^2 g'^2 a_x'}{3
   g^2}+\frac{2}{3} \gamma U f f'' a_x'-\frac{2 \gamma U f f' g' a_x'}{3 g}\nonumber\\&-&\frac{4}{3} \gamma U f''g A_t'h_{tx}
+\frac{4}{3} \gamma  U f' g'
   h_{tx} A_t'+\frac{4}{3} \gamma U f g'' h_{tx} A_t'-\frac{4 \gamma U fg'^2  A_t' h_{tx}}{3 g}+2 \gamma U fg' A_t' h_{tx}'\nonumber\\
   &+&2\gamma U f g A_t'h_{tx} +\ldots\,,
\end{eqnarray}
which one can check that it agrees with  the $U(1)$ current in the dual field theory.\footnote{As in the EMDA theory, we assume that the couplings $Z(\phi)$ and $U(\phi)$ are both finite at the boundary, so that the terms with the gauge field are always finite in the UV, and additional counter-terms are not needed.}
\begin{eqnarray}\label{ecurrent}
\left\langle J^x\right\rangle=\frac{\delta S}{\delta A_x}\Big|_{r\rightarrow r_{boundary}}=-\sqrt{-g}\left(ZF^{rx}-4\gamma U C^{rx\mu\nu}F_{\mu\nu}\right)|_{r\rightarrow r_{boundary}}.
\end{eqnarray}
To obtain the heat current, we need to find another radially conserved current.  For general gravity theories, this current has already been constructed in \cite{pope:1708} which is similar as Wald's procedure:\begin{eqnarray}\label{accurrent}
J^Q=2\sqrt{-g}\left(\frac{\partial L}{\partial R_{rx\rho \sigma}}\nabla_\rho \xi_\sigma+2\xi_\rho\nabla_\sigma \frac{\partial L}{\partial R_{rx\rho \sigma}} \right)-\xi^\rho A_\rho J^E\,,
\end{eqnarray}
where $\xi=\partial_t$ is the time-like Killing vector. On the other hand, the Weyl correction can also be re-expressed as
\begin{eqnarray}\label{Weylex}
C_{\mu \nu \rho \sigma}F^{\mu\nu}F^{\rho\sigma}=R_{\mu \nu \rho \sigma}F^{\mu\nu}F^{\rho\sigma}-2R_{\mu \nu}{F^{\mu}}_{\rho}F^{\nu\rho}+\frac{1}{3}RF^2\,.
\end{eqnarray}
 Then we obtain that
\begin{eqnarray}\label{hcurrent}
J^Q&=&f g h_{tx}'-f'g
   h_{tx}-\frac{2 \gamma U f^2 g' A_t' a_x'}{g}-2 \gamma U' f^2 A_t'a_x'-2 \gamma U  f^2A_t'' a_x'-2 \gamma U f^2  A_t'a_x''\nonumber\\
   &+&\frac{2}{3} \gamma U f'g h_{tx}A_t'^2-\frac{8}{3} \gamma U f g' h_{tx} A_t'^2-2
   \gamma U' f g  A_t'^2h_{tx}-4 \gamma U  f g  A_t'
   A_t''h_{tx}\nonumber\\
   &-&\frac{2}{3} \gamma U f g A_t'^2 h_{tx}'+\ldots\,,
\end{eqnarray}
which one can check that the Weyl corrections are vanishing at the infinite boundary and $J^Q$ equals the heat current in the dual field theory \cite{donos:2014}. The radial conservation of $J^E$ and $J^Q$ allow us to express them in terms of the horizon data. Near the horizon,
the $x-x$ and $r-x$ components of the Einstein equation reduce into respectively
\begin{eqnarray}
&&
\label{xxe}
\frac{1}{4} f g\phi
   '^2+\frac{g V}{2}-\frac{1}{3} \gamma  g U f'' A_t'^2+\frac{2}{3} \gamma  U f' g'A_t'^2+\frac{2}{3} \gamma  g U f' A_t' A_t''-\frac{1}{4} g ZA_t'^2+\frac{g f''}{2}+\frac{f' g'}{2}=0\,,
   \nonumber
   \\
   &&
   \
   \\
&&
-\frac{\gamma E_x U A_t' f''}{3 f}-\frac{2 \gamma  E_x
   U A_t' f' g'}{3 f g}+\frac{E_x Z A_t'}{2f}-\frac{1}{3} \gamma  g h_{rx} U A_t'^2 f''+\frac{2}{3}
   \gamma  h_{rx} U A_t'^2 f' g'\nonumber \\
 &&-\frac{4 \gamma  \zeta  U A_t'^2 f'}{3 f}-\frac{1}{4} g h_{rx} Z
   A_t'^2+\frac{2}{3} \gamma  g h_{rx} U A_t'  A_t'' f'+\frac{1}{4} f{\phi'}^2 g h_{rx}
+\frac{1}{2}g h_{rx} f''\nonumber\\
   &&
   \label{rxe}
   +\frac{1}{2} h_{rx} f' g'+\frac{\zeta  f'}{2 f}+\frac{1}{2} g h_{rx} V+\frac{1}{2}k^2W h_{rx}=0\,.
\end{eqnarray}
Using (\ref{xxe}) to eliminate $V$ and $\phi'$ in (\ref{rxe}), it turns out that $h_{rx}$ behaves like
\begin{eqnarray}\label{hrxhorizon}
h_{rx}(r)&&= \frac{E_x}{f}\left(-\frac{Z A_t'}{k^2 W}+\frac{2 \gamma U f'' A_t'}{3 k^2 W}+\frac{4 \gamma U f' g' A_t'}{3
   g k^2 W}\right)\nonumber\\
   &&+\frac{\zeta}{f}\left(-\frac{f'}{k^2 W}+\frac{8 \gamma  U f'A_t'^2}{3 k^2 W}\right)_{r=r_h}+\ldots\,.
\end{eqnarray}
The regular conditions at the horizon should be chosen as follows
\begin{eqnarray}\label{IRcod}
a_x'&=& -\frac{E_x}{f}+\ldots\,,\\
h_{tx}&=& f h_{rx}+\ldots\,.
\end{eqnarray}
Plugging this back to (\ref{conservedcurrent1}) and (\ref{hcurrent}) and  using the following horizon formulas
\begin{eqnarray}\label{DC}
\sigma=\frac{\partial J^E(r_h)}{\partial E_x}, \ \ \bar{\alpha}=\frac{1}{T}\frac{\partial J^Q(r_h)}{\partial E_x}, \ \ \alpha=\frac{1}{T}\frac{\partial J^E(r_h)}{\partial \zeta},\ \ \bar{\kappa}=\frac{1}{T}\frac{\partial J^Q(r_h)}{\partial \zeta},
\end{eqnarray}
the DC conductivities can be expressed in terms of the horizon data as follows
\begin{eqnarray}\label{DC2}
\sigma&\approx&\left[Z+\frac{gZ^2 A_t'^2}{k^2 W}+\gamma U\left(\frac{2 g f'' ZA_t'^2}{3 k^2 W}-\frac{8 f' g'ZA_t'^2}{3 k^2 W}-\frac{2}{3}
     f''+\frac{2 f' g'}{3 g}\right)\right]_{r=r_h}\,,\\
\bar{\alpha}&\approx&\left[\frac{4 \pi
    g Z A_t'}{k^2 W}-\gamma U \left(\frac{8 \pi  gf'' A_t'}{3 k^2 W}+\frac{16 \pi f' g' A_t'}{3 k^2
   W}+\frac{8 \pi  g Z A_t'^3}{3 k^2 W}+8 \pi A_t'\right)\right]_{r=r_h} \label{for1}\,,\\
\alpha&\approx&\left[\frac{4 \pi  g Z
   A_t'}{k^2 W}+\gamma U\left(\frac{16 \pi   g f'' A_t'}{3 k^2 W}-\frac{16 \pi  f' g' A_t'}{3 k^2
   W}-\frac{32 \pi   g  Z A_t'^3}{3 k^2 W}\right)\right]_{r=r_h}\label{for2}\,,\\
\bar{\kappa}&\approx&\left[\frac{4 \pi  g f'}{k^2 W}-\gamma U\frac{40 \pi    g  f' A_t'^2}{3 k^2 W}\right]_{r=r_h}\,.
\end{eqnarray}
If we use the horizon relation:
\begin{eqnarray}\label{hori1133}
k^2 W-g f''+g ZA_t'^2+2 \gamma U g f'' A_t'^2-\frac{10}{3} \gamma  U f' g'
   A_t'^2-\frac{8}{3} \gamma U  g f' A_t' A_t''=0\,,
   \end{eqnarray}
to eliminate $f''$ and set $\gamma=0$, they reduce to the results in the EMDA theory \cite{donos:2014}. The thermal conductivity in the open circuit condition is defined by
\begin{eqnarray}\label{DC2}
\kappa=\bar{\kappa}-\frac{\bar{\alpha}\alpha T}{\sigma}\,.
\end{eqnarray}
Eliminating $k^2W$ by using (\ref{hori1133}), it is finally obtained as
\begin{eqnarray}\label{kappaap}
\kappa\approx \left(\frac{4 \pi  f'}{f''}-\frac{8 \pi
   \gamma  U f' A_t'^2}{3 f''}-\frac{32 \pi  \gamma  U {f'}^2 A_t' A_t''}{3 {f''}^2}\right)_{r=r_h}+O(\gamma^2)\,.
\end{eqnarray}
At low temperatures, the last term with $f'^2|_{r=r_h}$ can be neglected. Then it agrees with (\ref{Tconducitivity}) in the main text.

Furthermore, from (\ref{for1}) and (\ref{for2}), we find that $\alpha-\bar{\alpha}=16 \pi \gamma U A_t'\neq 0$, which implies that the time-reversal symmetry is broken according to the Onsager relation. This feature should be attributed to the introduction of the Weyl term since the difference between $\alpha$ and $\bar \alpha$ is $\mathcal O (\gamma)$. We leave the detailed analysis for future investigation.

\section{Butterfly velocity with Weyl corrections}\label{appxshock}
The butterfly velocity characterizes the propagation of information in a chaotic quantum system and can be measured through the out-of-time correlator(OTOC):
\begin{equation}
\langle[W(x^i,t_W),V(0,0)]\rangle_\beta\sim e^{\lambda_L (t_W-t^*-|x^i|/v_B)}\,,
\end{equation}
where $W$ and $V$ are two generic local Hermitian operators, $\lambda_L$ is the Lyapunov exponent, $t^*$ is the scrambling time and $v_B$ is the butterfly velocity.
In holography, the OTOC has been widely calculated in many gravity theories by solving a shockwave solution in a two-sided black hole \cite{blake:1603,blake:1604,shenker:2013,malda:2015,lucas:2016,baggioli:2016,baggioli:2017,blake:1705,blake:1611,kim:1704,kim:1708,liu:2017,Mokhtari:2017,
Shenker:1412,swingle:1603,Ling:16101,Ling:16102,Alishahiha:1610,Cai:1704,Wu:1702,Wu:1706,Ling:1707,Qaemmaqami:1705,Qaemmaqami:1707,Giataganas:1708,lucas:1710,Huang:2017ohr,Mezei:1612,Jahnke:1708}.

For simplicity, we rewrite the Einstein equation (\ref{EE}) into
\begin{equation}\label{EEK}
G_{\mu\nu}-\gamma U\Big(G_{1\mu\nu}+G_{2\mu\nu}+G_{3\mu\nu}\Big)=T_{\mu\nu},
\end{equation}
where $T_{\mu\nu}$ is the stress tensor. In Kruskal coordinates, the black hole solution (\ref{newmetric}) can be re-expressed as
\begin{eqnarray}
ds^2\,=\,2\,A(uv)\,du\,dv\,+\,B(uv)\,dx^i\,dx^i, \nonumber\\
A_{\mu}=(-C(uv) v, C(u v) u,0,0), \quad\chi^I=k \delta_i^{I} x^i\,.
\end{eqnarray}
The horizon location $r=r_h$ in the original coordinates now is $uv=0$. And the Kruskal coordinates are defined by:
\begin{eqnarray}\label{d}
u\,v\,=\,-\,e^{f'(r_h)\,r_*}, \ \ \ u/v\,=\,-\,e^{-\,f'(r_h)\,t},
\end{eqnarray}
where $dr_*=\frac{dr}{f(r)}$. Moreover the functions appearing in the metric are related by the following relations:
\begin{eqnarray}\label{e}
A(uv)\,=\,\,\frac{2}{uv}\,\frac{f(r)}{f'(r_h)^2}, \ \ B(uv)\,=g(r), \ \  C(uv)=\frac{1}{u v}\frac{A_t(r)}{f'(r_h)}.
\end{eqnarray}
where $f(r)$, $g(r)$ and $A_t(r)$ are the metric components and gauge field in the original coordionates. We perturb the spacetime with an operator at $x^i=0$\footnote{Since the system is isotropic, we will omit the spatial index $i$ from now on.} and $t_L=t_W$, \textit{i.e.} a localized shock-wave; the butterfly velocity corresponds to the rate of growth of this perturbation.

The localized stress tensor of such a perturbation is given by:
\begin{equation}
T_{uu}^{shock}\,=\,E_0\,e^{2\,\pi T\,t_W}\,\delta(u)\,a(x).
\end{equation}
Then for large distance $|x|\gg1$, one can  replace $a(x)$ with a delta function  approximately. The shockwave solution corresponds to the geometry where there is a shift $v\rightarrow v\,+\,h(x,t_W)$ once one crosses the horizon $u=0$. The backreaction produces a perturbation in the spacetime metric of the form:
\begin{equation}
ds^2\,=\,2\,A(uv)\,du\,dv\,+\,B(uv)\,dx^i\,dx^i\,-\,2\,A(uv)\,h(x,t_W)\,\delta(u)\,du^2\,,
\end{equation}
and the stress tensor should get modified as \cite{swingle:1603,stensor:1994}:
\begin{equation}
\delta T_{uu}=T_{uu}^{shock}\,-\,2\,h(x,t_W)\,\delta(u)\,T_{uv}^0\,,
\end{equation}
where the second term is the leading contribution from the deformed geometry. Then the first order Einstein equation becomes\footnote{Note that our results (\ref{c4}) and (\ref{c5}) disagree with that in \cite{Mokhtari:2017}. We guess that the Weyl  corrections in the Einstein equation was missed in that paper.}
\begin{eqnarray}\label{c4}
\left(\partial_i^2-m^2\right)h(x^i,t_w)=\frac{3A(0)B(0)\,E_0 e^{2\pi T t_w}\delta(x)}{3A(0)^2+16\gamma U(0) C(0)^2}\,,
\end{eqnarray}
where the effective mass reads:
\begin{eqnarray}\label{c5}
&&
m^2=\frac{3 A(0)^3B'(0)-8 \gamma U(0)C(0)\left[A(0)B'(0)C(0)+4A(0)B(0)C'(0)-2A'(0)B(0)C(0)\right]}{3A(0)^4+16\gamma U(0) A(0)^2C(0)^2}\,,
\nonumber
\\
\end{eqnarray}
Solving the equation, we find that at large distances the solution takes the form:
\begin{equation}
h(x,t_W)\,\sim\,\frac{E_0\,e^{2\,\pi T (t_W\,-\,t^*)\,-\,m\,|x|}}{|x|^{1/2}}\,,
\end{equation}
where $t^*\sim \frac{1}{\lambda_L} \text{Log}\frac{1}{G}$ is the scrambling time.

As is pointed in \cite{Roberts:1409},  the profile of the shockwave, $h(x,t_W)$, corresponds to the OTOC of two generic local operators inserted at different locations and times with the spatial interval $x$ and temporal interval $t_W$. Then, the Lyapunov exponent and the butterfly velocity can be extracted as
\begin{equation}\label{lvb}
\lambda_L\,=\,2\,\pi T=\frac{1}{\tau_L}\,,\qquad v_B\,=\,\frac{2\,\pi T}{m}.
\end{equation}
The final step is to re-express $A(0)$, $B(0)$, $C(0)$ and their derivatives in the original $(t,r)$ coordinates.
Near the horizon we expand the quantities as follows
\begin{eqnarray}\label{d}
&&uv=-\kappa_0(r-r_h)+\ldots,\\
&&f(r)= f'(r_h) \,(r-r_h)+\frac{f''(r_h)}{2}(r-r_h)^2\ldots,\\
&&g(r)=g(r_h)+g'(r_h)(r-r_h)+\ldots,\\
&&A_t(r)=A_t'(r_h)(r-r_h)+\frac{A_t''(r_h)}{2}(r-r_h)^2+\ldots,
\end{eqnarray}
where $\kappa_0$ is a positive constant whose value is not important. On top of this, we have
\begin{eqnarray}\label{e}
&&A(0)=-\frac{2}{\kappa_0 f'(r_h)}+\ldots, \ \ A'(0)=\frac{f''(r_h)}{\kappa_0^2 f'(r_h)^2}+\ldots,\\
&&B(0)=g(r_h),\ \ B'(0)= -\frac{g'(r_h)}{\kappa_0}+\ldots, \\
&&C(0)=-\frac{A_t'(r_h)}{\kappa_0 f'(r_h)}+\ldots, \ \ C'(0)=\frac{A_t''(r_h)}{2 \kappa_0^2 f'(r_h)}+\ldots\ .
\end{eqnarray}
Then (\ref{c5}) can be re-expressed as
\begin{equation}\label{m2final}
m^2\,=\frac{3 f'(r_h) g'(r_h)-2\gamma U(r_h) A_t'(r_h)\Big(2 f'(r_h) g(r_h)A_t''(r_h)+f'(r_h) g'(r_h)A_t'(r_h)-f''(r_h)g(r_h)A_t'(r_h)\Big)}{6+8\gamma U(r_h)A_t'(r_h)^2}\,.
\end{equation}
This is the result for general values of $\gamma$. In this work, we focus on the physics at both of small $\gamma$ and low temperature limits. However, the final result may depend on which limit we take first. If we take the small $\gamma$ limit first, we obtain that
\begin{eqnarray}\label{butterflyv}
v_B^2&\approx&\frac{f'(r_h)}{2 g'(r_h)}-\frac{\gamma U(r_h) g(r_h) A_t'(r_h)^2 f''(r_h)}{3 {g'(r_h)}^2}+\frac{\gamma U(r_h) {A_t'(r_h)}^2
   f'(r_h)}{g'(r_h)}\nonumber\\
   &+&\frac{2 \gamma  U(r_h)g(r_h) A_t'(r_h) A_t''(r_h) f'(r_h)}{3
   g'(r_h)^2}.
\end{eqnarray}
At low temperatures, $v_B\sim T^\beta$ ($\beta$ is a constant) plus some small Weyl corrections. While in the $\text{AdS}_2 \times \text{R}^2$ case, $f''(r_h)\gg f'(r_h)$ at the low temperatures. Then, if we perform $T\sim f'(r_h)\rightarrow 0$ before taking the small $\gamma$ limit, the last term in (\ref{m2final}) dominates, which just gives
\begin{equation}
m^2\,=\frac{2\gamma U(r_h)f''(r_h)g(r_h)A_t'(r_h)^2}{6+8\gamma U(r_h) A_t'(r_h)^2}.
\end{equation}
In this case, we have
\begin{eqnarray}
v_B^2\approx \frac{3{f'(r_h)}^2}{4\gamma U(r_h)f''(r_h)g(r_h)A_t'(r_h)^2}\sim\frac{T^2}{\gamma\mu^2}.
\end{eqnarray}
which implies that $v_B$ is much slower than $v_B\sim \sqrt{T}$ for $\gamma \gg \frac{T}{\mu}$. Therefore, the result of $v_B$ highly depends on the order of manipulating the two limits.
\end{appendix}

\end{CJK*}
\end{document}